\documentstyle[preprint,aps]{revtex}
\begin{document}
\preprint{IISc-CTS-5/96}

\title{Universal Correlations in Random Matrices:\\
Quantum Chaos, the $1/r^2$ Integrable Model, and Quantum Gravity}

\author{
Sanjay Jain \cite{email}}

\address{Centre for Theoretical Studies \\
Indian Institute of Science, Bangalore-560012, India\\
and\\
Jawaharlal Nehru Centre for Advanced Scientific Research\\
Jakkur, Bangalore 560064, India}

\maketitle

\begin{abstract}
Random matrix theory (RMT) provides a common mathematical formulation of
distinct physical questions in three different areas:
quantum chaos, the 1-d integrable model with the $1/r^2$
interaction (the Calogero-Sutherland-Moser system), and 2-d
quantum gravity. We review the connection of RMT with these areas.
We also discuss the method of loop equations for determining correlation
functions in RMT, and smoothed
global eigenvalue correlators in the 2-matrix model for gaussian
orthogonal, unitary and symplectic ensembles.
\end{abstract}

\vskip 0.5cm
\noindent {\bf 1. Introduction}\hfill\break
In the last few years it has been realized that matrix models form a
bridge between two apparently distinct subjects: the Calogero-Sutherland-Moser
integrable model of particles in 1-d interacting via the $1/r^2$ potential (CSM
model),
and quantum chaos. Correlations of the eigenvalue
density in quantum chaotic systems and spacetime dependent
particle density correlators in the CSM model both map onto
the same eigenvalue correlator in a random matrix model.
Furthermore very
closely related correlation functions in the matrix model
appear in the context of QCD, string theory and 2-d
quantum gravity - the correlation functions of `loop operators',
which have a geometric
interpretation in terms of surfaces. Thus different physical questions
in these three vastly different areas reduce to the same mathematical
question formulated in terms of matrix models. In this review we
discuss these questions and their common mathematical formulation.
We also review the method of loop equations,
a technique for calculating these
correlators that originated in the QCD/string theory context,
and its application
to obtain correlation functions in the $2$-matrix
model that are relevant for quantum chaos and the CSM model.

In section 2 an introduction to RMT is given.
We collect some recent results for correlation functions in 1- and 2-matrix
models that are relevant to the sequel. Sections 3
and 4 describe the application of RMT in the field of quantum chaos and
for obtaining the static correlators of the CSM model,
respectively. Section 5 discusses a deeper relationship between quantum chaos
and the CSM model, that is, the equivalence of parametric correlations
in quantum chaos and time dependent correlations in the CSM model via a mapping
onto a 2-matrix model.  Two-dimensional
quantum gravity and its relationship with matrix models is described in
section 6.
Section 7 discusses loop equations for determining correlation
functions in matrix models and derives some of the results for
eigenvalue correlation functions mentioned in section 2. Section 8 contains
a brief summary.

\vskip 0.5cm
\noindent {\bf 2. Random Matrix Models}\hfill\break
Consider the partition function
\begin{equation}
Z =\int dA~~ e^{-S(A)},\label{eq:Z}
\end{equation}
where $A$ is an $N\times N$ hermitian matrix, $dA$ is the standard
$U(N)$ invariant measure for hermitian matrices, and $S(A)$ is a
$U(N)$ invariant action of the form
\begin{equation}
S(A) = N {\rm Tr} V(A), \quad\quad V(A) = \sum_{n=1}^\infty {1 \over n}
g_n A^n.\label{eq:S}
\end{equation}
This partition function defines the normalization of a probability
distribution for the matrix $A$, namely $P(A) = (1/Z)e^{-S(A)}$, in
which the expectation value of any function $f(A)$ is given by
$\langle f \rangle = \int \, dA \, P(A) \, f(A)$. Typical examples of
functions whose expectation values and correlation functions
will be of interest are the Green's function or
`loop operator' $\hat{W}(z) \equiv  (1/N) {\rm Tr} (z-A)^{-1}$, and the
eigenvalue density operator $\hat{\rho}(x) \equiv (1/N) {\rm Tr}\; \delta
(x-A)$.
One is mostly interested in this matrix model in the large $N$ limit.

The $U(N)$ symmetry of the above probability distribution (namely symmetry
under the transformation $A \rightarrow UAU^{\dag}$, with $U$ any $N\times N$
unitary matrix) can be used to project it down to a joint probability
distribution of the eigenvalues $\lambda_1,...,\lambda_N$ of $A$.
This probability distribution is given by
\begin{equation}
P(\lambda_1,...,\lambda_N) = N_0 \prod_{i<j}\, {|\lambda_i - \lambda_j|}^\beta
\,
e^{-N\,\sum_{i=1}^N\,V(\lambda_i)}, \label{eq:P}
\end{equation}
where $\beta = 2$. The projection is achieved by using the symmetry to
diagonalize
$A$ : $A = UDU^{\dag}$, $D =$ diag$(\lambda_1,...,\lambda_N)$. The `angular
variables' $U$ decouple from the eigenvalue integration because of the $U(N)$
symmetry and their integration provides merely an overall
factor absorbed in the normalization constant $N_0$. The Van der Monde
determinant factor
$\prod_{i<j}\, {|\lambda_i - \lambda_j|}^\beta$
is simply the jacobian of the change of variables from $A$
to $D$ and $U$.

As a consequence of this projection the expectation value of any `observable'
(any $U(N)$ invariant function $f(A)$; in particular $\hat{W}(z)$ and
$\hat{\rho}(x)$ above) is given by
\begin{equation}
\langle f \rangle = N_0 \int \, d\lambda_1...d\lambda_N\,
e^{-E(\lambda_1,...,\lambda_N)} \, f(D), \label{eq:EXPECT}
\end{equation}
with
\begin{equation}
E(\lambda_1,...,\lambda_N)= N\,\sum_{i=1}^N\,V(\lambda_i) -
\beta \sum_{i<j} \log {|\lambda_i - \lambda_j|}. \label{eq:E}
\end{equation}
This provides a very useful physical picture of the matrix model: the
eigenvalue $\lambda_i$ can be thought of as the position coordinate
of the $i$th particle in a 1-dimensional gas of $N$ particles
(the Dyson gas \cite{DYSONi}) in which
all particles experience the same external potential $NV(x)$,
and repel each other logarithmically, giving rise to a (static)
energy function $E$. $\langle f \rangle$ is merely the Boltzmann
weighted average of the function $f(D)$ of the position variables.

If $A$ belonged to the ensemble of $N \times N$ real symmetric matrices
rather than hermitian matrices (that is to say, for real symmetric $A$,
$P(A)$ is the same as that given above, but zero otherwise), equations
(\ref{eq:P}-\ref{eq:E})\ remain unchanged, except that now $\beta = 1$.
Similarly, if $A$ belonged to the ensemble of $N \times N$ real self-dual
quaternions, then $\beta = 4$. The three ensembles corresponding to
$\beta = 1,2,4$ are referred to as the orthogonal, unitary and symplectic
ensembles respectively, corresponding to the symmetry group in each case
For details, see, e.g., \cite{MEHTA}. $\beta$, which determines the
strength of the eigenvalue repulsion, simply measures the unlikelyhood
of finding a matrix in the ensemble with a pair of equal eigenvalues.

The probability distribution (\ref{eq:P}) depends upon the parameters
$g_n$ of the potential $V$. The simplest
cases are the gaussian ensembles (where $g_2 \equiv \mu > 0,\;\; g_1 =
g_n = 0$ for $n \geq 3$)
for which the one-point function (or eigenvalue density)
$\rho(x) \equiv \langle \hat{\rho}(x) \rangle$
obeys the semicircle law \cite{WIGNERi}
\begin{equation}
\rho(x) ={2\over {\pi a^2}}\sqrt{a^2-x^2},~~ |x|\leq a,
\quad\quad {\rm and}\quad \quad \rho(x) = 0,~~ |x| > a,\label{eq:DENSITY}
\end{equation}
with $a^2=2\beta/\mu$.
In general, when $V(x)$ has a single minimum,
$\rho(x)$  is nonvanishing on a single segment $[a,b]$ of the real line
and has the form
$\rho(x) = P(x)\sqrt{(x-a)(b-x)}$, where $P(x)$ is
a polynomial in $x$ whose degree depends upon the degree of $V$. Both
$a$ and $b$ as well as the coefficients of $P(x)$ depend upon the $g_n$,
thereby making $\rho(x)$ nonuniversal.

However, it turns out that higher correlation functions
of the eigenvalue density and the loop operator
exhibit a remarkable universality (independence
of the choice of $g_n$) in certain domains. For example, the exact
two point density-density correlator
$\langle\hat{\rho}(x) \hat{\rho}(y)\rangle$ for an arbitrary even
polynomial potential $V$ for the unitary ensemble ($\beta=2$),
computed in \cite{BZi}, though not universal in general,
is universal in two physically interesting domains.
The first domain is the `local' region where the interval $[x,y]$
contains a finite number ($O(N^0)$) of eigenvalues and is located
a finite distance from the endpoints. Then, the result, for arbitrary $V$, is
\begin{equation}
{{\langle \hat{\rho}(x) \hat{\rho}(y) \rangle} \over
{\rho(x)\rho(y)}}
= 1-{{\sin^2 \delta} \over {\delta^2}}, \quad\quad \delta \equiv
\pi N (x-y) \rho({{x+y} \over 2}),
\label{eq:FINE}
\end{equation}
which is universal when expressed in terms of variables that are rescaled
by the factor $\rho({{x+y} \over 2})$ that determines the local mean level
spacing. This is a `fine grained' correlator that oscillates over
distances of the order of mean eigenvalue spacing,
and has a nonsingular limit as $x-y \rightarrow 0$. (\ref{eq:FINE}) was
derived originally for the circular unitary ensemble (CUE: ensemble of
unitary matrices, whose eigenvalues lie on a circle) in \cite{DYSONii}\
and for the gaussian unitary ensemble (GUE: ensemble of hermitian matrices
with a quadratic $V$) in \cite{WIGNERii}. Its universality for the
unitary ensemble ($\beta=2$) was extended
to all $V$'s that correspond to the classical orthogonal polynomials
in \cite{FK}\ and to all even polynomial $V$'s in \cite{BZi}.
Expressions analogous to (\ref{eq:FINE}) exist for the other two
ensembles ($\beta = 1,4$) \cite{DM,MEHTA}. The universality of all these local
fine grained correlators with respect to arbitrary potentials has
been shown in \cite{HW}. For other extensions of the universality of
(\ref{eq:FINE}) and the higher point local correlators see \cite{UNIV}.

While the universality of local correlators has long been expected
on empirical grounds (see next section), it is somewhat more of a
surprise that even certain `global' correlators are universal.
This is the domain where $[x,y]$ contains a large number of eigenvalues
including upto $O(N)$, but the correlator is `smoothed' by independently
averaging over $x$ and $y$ over intervals much greater than the mean
eigenvalue spacing. In this domain the result for the connected two point
function in the large $N$ limit is
\begin{equation}
\langle \hat{\rho}(x) \hat{\rho}(y) \rangle_c =
-{1 \over {\pi^2 N^2 \beta}}{1 \over {(x-y)^2}}
{{a^2-xy} \over {[(a^2-x^2)(a^2-y^2)]^{1/2}}}.
\label{eq:TWOPOINTi}
\end{equation}
This is universal in that it depends on the $g_n$ only through the endpoint
of the spectrum $a$. (For even potentials the endpoints are $\pm a$. For
non-even
potentials, the result is still (\ref{eq:TWOPOINTi}), but now $a$ represents
half the width of the support, and $x,y$ on the r.h.s. of (\ref{eq:TWOPOINTi})
are
shifted by the midpoint of the support.) This is
a `global' correlator in the sense that $x$ and $y$ could be
anywhere in the interval $[-a,a]$, close to the endpoints or far away
(provided only that $x-y$ contains a large enough number of eigenvalues
for `smoothing').
(\ref{eq:TWOPOINTi}) was derived originally for
the gaussian ensembles in ref. \cite{FMP}. Its universality with respect to
arbitrary
potentials for $\beta=2$ follows from the results of ref. \cite{AMJ}
obtained for the two-point function of loop operators.
In \cite{BZi}\ (\ref{eq:TWOPOINTi})
was shown to be the smoothed version of the exact two-point density-density
correlator
for arbitrary even polynomial potentials in the unitary ensemble.
(\ref{eq:TWOPOINTi}) has been proven to be universal for arbitrary potentials
$V$ which give rise to an eigenvalue density with support in a single segment
for all three ensembles \cite{BEENi}. For these potentials the three and
higher point smoothed correlators vanish to leading order in $1/N$
\cite{BZi,BEENi,POLITZER}\
and are in general nonuniversal in higher orders \cite{EYNARD}.

There is yet another type of universality of smoothed
correlators at multicritical points \cite{KAZAKOV}\ that arises
in the double scaling limit \cite{DS}\ studied in the context of
2-d gravity and noncritical string theory \cite{BW}.
Here only the region close
to the endpoint of the support of $\rho(x)$ matters, and correlations
depend on the $g_n$ only through the number of zeroes of the polynomial $P(x)$
that coincide with the endpoint $a$ (the level of multicriticality)
\cite{AWNBB}. Further, there is evidence of a universality of finegrained
correlators at the spectrum edge even in the absence of a double scaling
limit \cite{EDGE}.

The 2-matrix model is defined by the partition function
\begin{equation}
Z =\int dA~~ dB~~ e^{-S}\label{eq:Zii}
\end{equation}
where $A$ and $B$ are $N \times N$ matrices drawn from any of the
above mentioned ensembles, and we will restrict the discussion to $S$ of
the type $ S=N {\rm Tr}~(V_1 (A) +V_2(B)-c AB)$.
When $A$ and $B$ are drawn from different
ensembles, e.g., $A$ is drawn from the gaussian
orthogonal or symplectic ensemble ($\beta=1$ or $4$) and $B$
from the gaussian unitary ensemble ($\beta=2$), the eigenvalue correlations
(relevant for parametrizing time reversal symmetry breaking in
chaotic systems) have been computed in
\cite{TRANSITION}.
Here we will be primarily interested in the case where
both $A$ and $B$ are drawn from the same ensemble.

The smoothed, global expression for the 2-point function
$\rho_{AB} (x,y)\equiv
\langle \hat{\rho}_A (x)\hat{\rho}_B (y)\rangle_c
\equiv \langle \hat{\rho}_A (x)\hat{\rho}_B (y)\rangle -
\langle \hat{\rho}_A (x)\rangle \langle\hat{\rho}_B (y)\rangle$,
(where $\hat{\rho}_A (x)\equiv {1\over N} {\rm Tr}\; \delta (x-A)$ and
similarly $\hat{\rho}_B(x)$) is, in the large $N$ limit,
\begin{eqnarray}
\rho_{AB}(x,y)
&=&-{1\over {2\pi^2 N^2}}{1\over {\beta a^2}}{1\over
{\cos\theta \cos\alpha}}\nonumber \\
&&[{{1+{\rm ch} u \cos(\theta+\alpha)}\over
{({\rm ch} u+\cos(\theta+\alpha))^2}}
+{{1-{\rm ch} u \cos(\theta-\alpha)}\over
{({\rm ch} u-\cos(\theta-\alpha))^2}}].\label{eq:D}
\end{eqnarray}
Here $V_1(x) = V_2(x) = (1/2)\mu x^2$, $u\equiv \ln ({\mu\over c})$,
$x \equiv a\sin\theta$, $y\equiv a\sin\alpha$. The one point function
$\rho(x) \equiv \langle \hat{\rho}_A(x) \rangle = \langle \hat{\rho}_B(x)
\rangle$
is still given by the semicircle law  except that the endpoint of its
support is now $a=({{2 \beta \mu} \over {\mu^2-c^2}})^{1/2}$,
and (\ref{eq:D}) is valid for $|x|,|y|< a$.
(\ref{eq:D}) was computed in \cite{BZii}\
for $\beta=2$ using a diagrammatic technique and
in \cite{DJS}\ for $\beta=1,2,4$ using the method of loop
equations \cite{LOOPEQi}.

The local fine-grained correlators in the 2-matrix model
for the gaussian ensembles were
computed in \cite{SA}\ ($\beta=1,2$) and \cite{SLAiii}\
($\beta=4$) using the the supersymmetry technique \cite{EFETOV}.
For $\beta=2$, global fine-grained correlations are known \cite{DBZ}\
(see also \cite{BR})
using the method of orthogonal polynomials \cite{OP}.
The local correlators near
the centre of the eigenvalue distribution have been shown to be universal
(independent of the form of the potential) in
\cite{HW}\ (see \cite{SSA}\ for numerical evidence and \cite{SHASTRY}\
for a physical argument). The smoothed global correlator (\ref{eq:D})
although derived for a quadratic potential reduces, in two different
limits, to expressions that are universal. The first is when
$x,y$ are restricted to the region near the centre of the cut; then
the expression (see section 5) is the local smoothed correlator
\cite{SLAi} \cite{BEENii} which is universal.
The second is when $\mu$ and $c$ are tuned such
that the probability distribution $e^{-S}$ has vanishing weight
over configurations in which $A \neq B$, in which case (\ref{eq:D})
reduces to the universal global 1-matrix result (\ref{eq:TWOPOINTi}).
However, the universality of (\ref{eq:D}) in general remains unproven.

Correlators in the double scaling limit of the $\beta=2$ 2-matrix model with
non-quadratic potentials have been studied in \cite{DSii,LOOPEQiii}

\vskip 0.5cm \noindent
{\bf 3. Quantum Chaos}\hfill\break
The subject of quantum chaos is at present defined as the study of the
quantum properties of a system that is classically chaotic.
Consider a classically chaotic system described by some hamiltonian
$H=H(p_1,q_1,p_2,q_2,...)$. A signature of classical chaos is that
trajectories are extremely sensitive to initial conditions.
The question is, what can one say about the quantum system?
Here we restrict ourselves to questions regarding the spectrum of the quantum
hamiltonian.
In general, the analytic calculation of the energy eigenvalues is a
hopeless task. However, there is a substantial amount of data on the
discrete eigenvalues $E_i$ obtained numerically for model systems and
experimentally
for real systems. From this data one obtains various statistical properties
of the spectrum. For example, one can in principle determine the correlation
functions of the density
of eigenvalues $\langle \rho(E_1)\rho(E_2)...\rho(E_n) \rangle$, where $\rho(E)
\equiv
\sum_i\delta (E - E_i)$. ($\langle\;\rangle$ is an average over many
energy levels of the hamiltonian in question keeping the differences $E_i -
E_j$ fixed.)
In practice one computes various statistical measures from the data that are
related to these correlation functions (for reviews see
\cite{REICHL,HAAKE,MEHTA,BETAL}), for example, the distribution of
level spacings,
(i.e., the probability for the spacing between consecutive energy levels to
lie in a specified range). The result of these empirical analyses
is that while the one-point function $\langle \rho(E) \rangle$ varies
from system to system, the local
two and higher point correlation functions as well as the
level spacing distribution, for sufficiently high energy levels
not related by any symmetry, exhibit a universal
(i.e., system independent) behaviour for a wide variety of systems.

This observed universality of local eigenvalue statistics suggests that there
might be a probability distribution
of energy eigenvalues with universal features characterizing generic chaotic
systems.
It was Wigner's insight \cite{WIGNERi}, in the context of nuclear
physics, that the statistical spectra of complex nuclei should be obtainable
by taking the extreme view that one knows essentially nothing about
the hamiltonian, that is, the hamiltonian is drawn completely at
random from the space of all possible hamiltonians. This led him to
propose gaussian random matrix ensembles for theoretically calculating the
eigenvalue statistics, wherein the random matrix $A$ of the
1-matrix model in the previous section
corresponds to a hamiltonian, and correlation functions of the
`observable' $\hat{\rho}(x)$ in the matrix ensembles are to be compared
with empirically obtained correlators of the eigenvalue density
$\rho(E)$ ($x$ is identified with $E$ upto a rescaling to be discussed later).
Dyson \cite{DYSONi}\ showed that the three different ensembles in
random matrix theory should correspond to hamiltonians in three different
universality classes identified by symmetry: $\beta = 2$ corresponds
to hamiltonians with no time reversal invariance; if $H$ has time
reversal invariance, then $\beta =1$, except if the system has no
rotational invariance and has odd spin, in which case $\beta=4$.
It was proposed that the same statistical hypothesis should
apply to complex atoms \cite{RP}\ and to electrons in disordered
cavities (mesoscopic systems) \cite{GE}.
Later, it was suggested \cite{ZBT}\ and numerically observed
\cite{MKCVG}\ that spectra of classically chaotic systems obey level
repulsion as in RMT, and a proposal was made in \cite{BGS}\ that
that local correlations in RMT should provide the eigenvalue
statistics for any hamiltonian system in the strongly chaotic regime.

These proposals have been tested and verified in an amazing diversity of
systems. The list includes
the empirical spectra of complex nuclei \cite{NUCLEI},
atoms \cite{RP,ATOMS}, molecules \cite{MOLECULES},
and microwave cavities \cite{MICROWAVE},
the numerically obtained spectra of chaotic billiards \cite{BGS}
and other chaotic systems \cite{OTHER}, and analytic calculations
from microscopic models of
electrons in weakly disordered mesoscopic systems \cite{EFETOV}.
The empirically observed universality
mimics the universality mentioned for RMT in
section 2, in that the eigenvalue density is nonuniversal but higher
point local correlators and the spacing distribution is universal.
For systems in which one can tune parameters
in the hamiltonian to go from the classically integrable to the chaotic
regime, agreement with RMT is seen in the strongly
chaotic regime, and the statistics departs from RMT in the integrable
or near integrable domain.

Recently, the smoothed global
correlators have also found applications.
An application of the universality of (\ref{eq:TWOPOINTi}) is
an argument \cite{BEENi}\ for the
universality of conductance fluctuations of weakly disordered mesoscopic
conductors using the transfer matrix formalism \cite{TM}. Other examples
include possible application to the spectrum of the Dirac operator
in strongly coupled QCD \cite{QCD} and to the quantum Hall effect.

It seems that while being classically chaotic is usually a good indication of
a system being `generic' enough for Wigner's statistical hypothesis to apply,
there are some exceptions to this rule. One class of exceptions are
conductors with
strong disorder. Classically, these systems represent chaotic billiards since
an electron can bounce repeatedly from randomly placed impurities. In practice
their spectrum disagrees with random matrix theory. In this case the departure
can be explained in terms of a new quantum phenomenon, localization, that
arises
due to strong disorder, which disallows quantum states that are analogues of
the classical ballistic chaotic trajectories \cite{AS}. Other exceptions
are discussed in \cite{EXCEPTIONS}.
Therefore, until a better understanding of the precise domain of
applicability of RMT is reached (see \cite{BPR}\ for efforts using
semiclassical methods \cite{GUTZ}), the agreement
of energy level statistics of classically chaotic systems with
random matrix theory remains a fairly ubiquitous, but essentially empirical,
fact.

\vskip 0.5cm \noindent
{\bf 4. The $1/r^2$ Integrable Model: Static Correlators}\hfill\break
Consider the system of $N$ fermions in one dimension with the hamiltonian
(the Calogero-Sutherland-Moser (CSM) system \cite{CALOGERO,SUTHi,MOSER})
\begin{equation}
H = \sum_i^N (p_i^2 + \omega^2 x_i^2) + {1 \over 2}\beta (\beta -
2)\sum_{i<j}{1 \over
{(x_i - x_j)^2}}. \label{eq:HCSM}
\end{equation}
The particles are in a harmonic potential and interact with each other via the
$1/r^2$ potential with a strength determined by $\beta$
($\beta=2$ corresponds to free fermions). This is an integrable
model with $N$ conserved quantitities.
Its exact ground state wave function is given \cite{SUTHi}\ by
\begin{equation}
\psi(x_i,...,x_N) = C e^{-\omega\sum_i x_i^2/2} \prod_{i<j} |x_i -x_j|^{\beta
\over 2}. \label{eq:PSI}
\end{equation}
Therefore the expectation value of any position dependent
operator $F(x_1,...,x_N)$ in the ground state is given by
\begin{equation}
\langle F \rangle = \int dx_1...dx_N \psi^* F \psi
= C^2 \int dx_1...dx_N \prod_{i<j} |x_i -x_j|^{\beta}e^{-\omega\sum_i x_i^2}
F(x_1,...,x_N). \label{eq:EXPECTii}
\end{equation}
This has the same form as the expectation value of an observable in the
Wigner-Dyson distribution (\ref{eq:EXPECT}) with $V(x)={1 \over N}\omega x^2$,
$C^2 = N_0$, at $\beta = 1$, $2$, and $4$, with the position coordinates
of the particles corresponding to the eigenvalues of the random matrix.
Thus the matrix model (\ref{eq:Z}) describes free fermions for $\beta=2$ and
interacting
fermions for $\beta=1,4$.
In particular, for $F$ we may choose the $n$-point function of the
density operator in the CSM model, $\hat{\rho}(x) \equiv \sum_i \delta(x-x_i)$.
Then, for example the connected {\it{spatial}} equal-time density correlator in
the
$1/r^2$ model, exactly matches the matrix model and hence the quantum chaos
{\it{eigenvalue}} density correlator:
$\langle \hat{\rho}(x) \hat{\rho}(y) \rangle|_{\rm CSM\,\,model}
= N\langle \hat{\rho}(x/\sqrt{N}) \hat{\rho}(y/\sqrt{N})
\rangle|_{\rm matrix\,\,model}$, with the r.h.s. given by (\ref{eq:FINE}) or
(\ref{eq:TWOPOINTi}).

\vskip 0.5cm \noindent
{\bf 5. Dynamical Correlators, Perturbed Chaotic Hamiltonians, and the
2-Matrix Model} \hfill\break
We now discuss a deeper connection between quantum chaos and the CSM model,
established through the intermediary of a 2-matrix model. Consider a chaotic
hamiltonian $H$ consisting of an unperturbed part proportional to $H_0$
and a perturbation proportional to $H_1$:
\begin{equation}
H=H_0 \cos \Omega \phi + H_1 {\sin \Omega\phi \over \Omega}.
\label{eq:HCHAOS}
\end{equation}
$\phi$ is a measure of the strength of the perturbation ($\Omega$
is a constant, introduced for convenience, to be specified shortly).
Given $H_0$ and $H_1$, the eigenvalues of $H$ are a function of $\phi$.
One can think of a 1-d gas of particles whose positions are the eigenvalues
and $\phi$ as the time on which they depend. An interesting fact is that
the time evolution of this gas ($\phi$ dependence of the positions of the
eigenvalues) is given by a classical hamiltonian which is closely related to
the
CSM hamiltonian with some additional degrees of freedom. This is known
as the Pechukas gas \cite{PECHUKAS}. We do not discuss this further;
instead we focus on the `quenched' averages that arise when the spectral
statistics of $H$ is discussed using matrix model ensembles.
In particular we discuss the $\phi$ dependent correlator
$\langle \rho_{H_0}(E_1)\rho_H(E_2)\rangle_c$ which is of interest in
quantum chaos, since it
measures the correlations between the eigenvalues of
$H_0$ and $H$ as a function of the strength of the perturbation
($\rho_H(E)\equiv \sum\delta(E-E_i)$ where $E_i$ are the eigenvalues of $H$).

To see how this correlator is connected to the 2-matrix model,
note that the probability $P(H,\phi)dH$ that the full hamiltonian
lies in the volume element
$dH$ located at $H$ (in the space of $N\times N$ matrices of the appropriate
class given by $\beta$) is given by
$P(H,\phi)dH = \int dH_0 P_0(H_0) P(H,\phi|H_0)dH$,
where $P_0(H_0)dH_0$ is the probability that the unperturbed hamiltonian lies
in the volume element $dH_0$ located at $H_0$, and $P(H,\phi|H_0)dH$ is the
conditional probability that full hamiltonian lies in the volume element $dH$
located at $H$ given a fixed unperturbed part $H_0$.
But now $P(H,\phi|H_0)dH=P_1(H_1)dH_1=const. \;P_1(H_1)dH$, where
$P_1(H_1)dH_1$ is the probability of the perturbation being in a volume element
$dH_1$
located at $H_1$,
and we have assumed that $H_1$ and $H$ come from the
same class of ensemble (same $\beta$), whereby, for fixed $H_0$, $dH_1=const.\;
dH$.
Then $P(H,\phi)dH = const. \int dH_0 P_0(H_0) P_1(H_1)dH$,
where $H_1$ is understood to be given in terms of $H_0$ and $H$ by
(\ref{eq:HCHAOS}). Thus the average of any $H$ dependent quantity will
be given by a 2-matrix integral (the two matrices to be integrated over
being $H_0$ and $H$), in which the factor $P_1(H_1)$, when expressed in terms
of $H_0$ and $H$, will involve ($\phi$ dependent) terms that couple
the two matrices together. In particular if we take both probability
distributions $P_0$ and $P_1$ to be gaussian:
$P_0(H_0) \sim {\rm exp}(-{\Omega^2 \over 2}{\rm Tr}H_0^2)$,
$P_1(H_1) \sim {\rm exp}(-{1 \over 2}{\rm Tr}H_1^2)$,
it follows that
\begin{equation}
P(H,\phi)dH = const. \int dH_0dH\; {\rm exp}(-{\rm Tr}[V(H_0)+V(H)-cH_0H]),
\label{eq:PROB}
\end{equation}
\begin{equation}
V(x)={1 \over 2}\mu x^2, \quad\quad
\mu={\Omega^2 \over \sin^2 \Omega\phi},\quad\quad
{\rm and} \quad\quad c={\Omega^2 \cos\Omega\phi \over \sin^2 \Omega\phi}.
\label{eq:MC}
\end{equation}

It is convenient to rescale $H_0= \sqrt{N} A$, $H=\sqrt{N} B$, and define
the 2-matrix model partition function (\ref{eq:Zii})
with $ S=N {\rm Tr}~(V (A) +V (B)-c AB)$.
{}From the symmetry between $H_0$ and $H$ in (\ref{eq:PROB})
it follows that the
two eigenvalue densities are equal, $\langle \hat{\rho}_A(x) \rangle
= \langle \hat{\rho}_B(x) \rangle \equiv \rho(x)$, and are
given by the semicircle law for gaussian $P_0$ and $P_1$.
The constant $\Omega$ is chosen to be $\Omega = \sqrt{\pi^2 \beta/2N}$,
which means that $\mu$ and $c$ are $O(1)$ when $\phi \sim O(1)$.
It is evident from the above that under the random matrix hypothesis
the desired 2-point correlator in quantum chaos is given by
\cite{TRANSITION,HAAKE,NS}
\begin{equation}
\langle \rho_{H_0}(E_1)\rho_H(E_2)\rangle_c|_{\rm quantum\,\,chaos}
= N\rho_{AB}(E_1^\prime,E_2^\prime), \label{eq:RHOCHAOS}
\end{equation}
with $E_i^\prime=N^{-{1 \over 2}}E_i$.

Altschuler, Simons, Szafer, and Lee provided evidence based
on numerical results for chaotic billiards and the Anderson model
for the universality of the l.h.s. of (\ref{eq:RHOCHAOS}) \cite{SSA,SA},
computed the matrix model local correlator for the r.h.s. of
(\ref{eq:RHOCHAOS})
and exhibited agreement with numerical results \cite{SA},
and conjectured that the r.h.s. of (\ref{eq:RHOCHAOS}) furthermore
equals the {\it dynamical} density-density
correlation function in the CSM model \cite{SLAi}, with the perturbation
strength $\phi$ in the quantum chaos problem playing the role of
time in the CSM model (and $E_i$ playing the role of the positions
$x_i$ of the CSM particles).

We now outline the proof of the latter conjecture \cite{NS}\
refering the reader
to \cite{SHASTRY} for a detailed review.
(See also \cite{HAAKE}\ for related observations and for a proof that holds
for smoothed correlators only, see \cite{BEENii}.) It is not difficult to see
that the probability distribution
$P(H,\phi|H_0)=const.\; exp(-{\rm Tr}[V(H_0)+V(H)-cH_0H])$
for the matrix elements of $H$ which appears in (\ref{eq:PROB})
is a solution of a Fokker-Planck equation describing
the independent Brownian motion of the matrix elements of $H$,
where the time $t$ is given by
\begin{equation}
t=-{{\log(\cos(\Omega\phi))} \over \Omega^2}, \label{eq:TIME}
\end{equation}
and the initial condition is that at $t=\phi=0$, the distribution
is localized on $H_0$, $P(H,0|H_0)=\delta(H-H_0)$. Dyson \cite{DYSONiv}\
had shown that when the elements of a matrix execute independent
Brownian motion, its eigenvalues execute a correlated Brownian motion
in which they repel each other by a logarithmic interaction, and derived
the corresponding Fokker-Planck equation for the time dependent
probability distribution of the eigenvalues. Sutherland \cite{SUTHii}\
showed that the eigenfunctions of this latter Fokker-Planck hamiltonian
were in one-to-one correspondence with the eigenfunctions of the
CSM hamiltonian. It is this correspondence, which extends the relationship
between the matrix model and the CSM model at the {\it ground state} level
(responsible for static correlators as discussed in the previous section)
to all the {\it excited states}, that is at the heart of the mapping between
the dynamical correlators of the CSM model and matrix models. One can
also map the CSM model to a continuous matrix model in which the matrix
$A$ is a function of a continuous variable $t$ \cite{SLAii}.
For the two-point function the precise mapping is \cite{NS}
\begin{equation}
\langle \rho(x,0) \rho(y,t) \rangle_c|_{\rm CSM\,model}
= N\rho_{AB}(x^\prime,y^\prime), \label{eq:RHOCSM}
\end{equation}
where $\hat{\rho}(x,t)=\sum_i\delta(x-\hat{x}_i(t))$, and $\hat{x}_i(t)$
is the time dependent position operator of the CSM model in the Heisenberg
picture.
The l.h.s. of (\ref{eq:RHOCSM}) as a function of
$x,y,t$, and parameters $\omega,\beta,N$ is thus expressed as a correlator
of the 2-matrix model (\ref{eq:Zii}) with couplings $\mu,c$ expressed in terms
of $t$ via eqs. (\ref{eq:MC},\ref{eq:TIME}), $\omega=\Omega^2/2$, and the
mapping works
for the values $\beta=1,2,4$ for which
the r.h.s. can be defined in terms of the appropriate matrix ensemble.

Eq. (\ref{eq:D}), substituted in (\ref{eq:RHOCSM}), therefore gives the
smoothed global spacetime dependent density-density correlator of the
CSM model. If one specializes (\ref{eq:D})
to the region near the centre of the semicircle ($x,y \ll a$),
and for $t \approx \phi^2 \ll O(N)$ one
recovers the previously known `translationally invariant' result
\cite{SLAi} \cite{BEENii}
\begin{equation}
\langle\hat{\rho}(x,0)\hat{\rho}(y,t)\rangle_c=-{2\over {\pi^2 \beta^2}}
{[(x-y)^2 - \pi^2\beta^2 t^2] \over {[(x-y)^2 + \pi^2\beta^2 t^2]^2}}.
\label{eq:RHOXY}
\end{equation}
This identifies the sound velocity as $v_s=\pi\beta$.
We mention that for the CSM model at other rational values of $\beta$
not covered by known matrix ensembles, local fine-grained dynamical correlators
have been obtained by the method of Jack Polynomials \cite{HA}.
See \cite{FORRESTER}\ for a proof of (\ref{eq:TWOPOINTi}) for
all even $\beta$.

Together with (\ref{eq:RHOCHAOS}) the equation (\ref{eq:RHOCSM}) establishes
a connection between perturbation strength dependent correlators in quantum
chaos
and dynamical correlators in the CSM model via the 2-matrix model. The r.h.s.
of (\ref{eq:RHOXY}) therefore also equals the parametric
eigenvalue correlator $\langle\rho_{H_0}(x)\rho_H(y)\rangle_c$
in the quantum chaos problem $H=H_0 + \phi H_1$, when $H_0$ and
$H_1$ belong to the same symmetry class.
We emphasize that the mapping from the quantum chaos problem to
the matrix model is an ansatz that has essentially empirical support.
On the other hand the mapping
between the matrix model and the CSM model discussed
above is an exact mathematical mapping.

\vskip 0.5cm \noindent
{\bf 6. Two Dimensional Quantum Gravity} \hfill\break
Matrix models have long been studied in quantum chromodynamics, where
the basic degree of freedom is a matrix valued field. The observation
\cite{THOOFT} that the $1/N$
expansion of SU($N$) QCD groups the Feynman diagrams of the theory
according to the topological classes of 2-dimensional surfaces
led to a vigorous study of the random matrix models in the large-$N$
limit \cite{BW}.
This has also led to matrix models being proposed as
models of 2-dimensional quantum gravity and string theory
\cite{GRAV}, where
one needs to sum over all 2-dimensional surfaces keeping track of
their topology.

A 2-dimensional surface can be latticized by triangulation,
which is a representation of the surface in terms of small, flat,
equilateral triangles (all of the same size) joined along their
edges. In this discrete representation, all information about the
local geometry of the surface is contained in the local `coordination number',
the number of triangles that meet at a vertex. The area of the
surface is equal to the total number of triangles (in units of the
area of each elementary triangle), the length of any boundary is
equal to the number of edges in that boundary, and the topology of the surface,
(given by its Euler characteristic $\chi \equiv 2-2h-b$ where $h$ is the number
of handles and $b$ the number of boundaries) is
$\chi = N_0 - N_1 + N_2$, where $N_0 =$ number of vertices, $N_1 =$
number of edges, and $N_2 =$ number of triangles in the triangulation.

In a path integral formulation of 2-dimensional quantum gravity one needs
to sum over all surfaces, assigning a specific weight to each surface.
This sum is defined to be the sum over all distinct triangulations
in the lattice version of the theory \cite{GRAV}. For example the partition
function
for pure gravity (without matter fields) is defined to be
\begin{equation}
Z(\kappa, \lambda) = \sum_{h=0}^\infty e^{\kappa(2-2h)}\sum_{A=4}^\infty
e^{-\lambda A} \tilde{Z}(h,A),\label{eq:Ziii}
\end{equation}
where $\tilde{Z}(h,A)$ is the number of distinct triangulations  of connected
oriented surfaces
with area $A$, number of handles $h$, and no boundary. $\kappa$ is referred
to as the gravitational constant since it multiplies the Einstein action
(which is proportional to $2-2h$ for closed surfaces in two dimensions),
and $\lambda$ the cosmological constant.

The above partition function can be obtained from the partition function
of a hermitian 1-matrix model:
$Z(\kappa, \lambda) = \log Z$ where the $Z$ on the r.h.s. is given by
(\ref{eq:Z}), with
$V(A) = (1/2)A^2 - gA^3$, $N=e^\kappa$
and $g=e^{-\lambda}$. This is apparent upon expanding (\ref{eq:Z})
as a power series in $g$, and collecting all the Feynman diagrams
which appear with the same power of $1/N$. Since every Feynman diagram
is a vacuum bubble made of propagators and cubic vertices only, its dual
diagram (obtained by joining
the centres of adjacent loops) is a triangulated
surface. Since the perturbative evaluation of the partition function
involves summing over all Feynman diagrams, one automatically gets
a sum over triangulations. The weight factor appearing with every
Feynman diagram containing $V$ vertices, $P$ propagators and $L$ loops
is $N^{V-P+L}g^V$, which becomes
$e^{\kappa(N_0-N_1+N_2) - \lambda N_2}$ as needed, since every dual
diagram has $N_0=L$ vertices, $N_1=P$ edges, and $N_2=V$ triangles.
The logarithm of $Z$ appears because in (\ref{eq:Ziii}) we need only
connected surfaces. The leading large $N$ contribution to $\log Z$
comes from closed connected surfaces with the spherical topology
($h=0$, $\chi=2$). The orthogonal and symplectic ensembles lead
to sums over non-orientable surfaces (see, e.g., \cite{UNORIENTED}).

Consider now the insertion of ${\rm Tr}\,A^l$ in the matrix model path
integral,
i.e., the quantity
$Z \langle {\rm Tr}\,A^l \rangle \equiv \int \, dA \, e^{-S} {\rm Tr} A^l$,
with $l$ a
positive integer. Expanding
the r.h.s. as a perturbation expansion in $g$, one notes that as before
all vertices of the Feynman diagrams are cubic, except one (corresponding to
the external insertion of ${\rm Tr} A^l$) where $l$ propagators meet. Thus all
the corresponding dual diagrams are triangulated surfaces as before,
but with one elementary plaquette which has $l$ sides. This distinguished
plaquette can be considered to be absent, i.e., the triangulated surface
can be thought of as having one boundary of length $l$. Thus the insertion
of ${\rm Tr} A^l$ in the path integral has the interpretation of inserting a
`loop' of length $l$ on the surface \cite{LOOPS}. Similarly the insertion of
${\rm Tr} A^l {\rm Tr} A^m$ can be seen to produce two loops of size $l$ and
$m$
on the surface. $\hat{W}(z) = (1/N) \sum_{n=0}^\infty
(1/z^{n+1}) {\rm Tr} A^n$ is called the loop operator because it is the
generating
function of loops of all lengths; $z$ is the fugacity for the length
of the loop. Thus correlation functions of the type
\begin{equation}
\langle\hat{W}(z_1)...\hat{W}(z_n)\rangle = {1 \over Z}\int dA \;
e^{-S}\hat{W}(z_1)...\hat{W}(z_n)\label{eq:LOOPCORR}
\end{equation}
are of direct physical significance in 2-d gravity and string theory
since they determine sums over surfaces with $n$ boundaries of
arbitrary length.

Now these correlation functions are closely related to the eigenvalue
correlations in the matrix model because of the simple identity
$\hat{\rho}(x) = (i/2\pi)\lim_{\epsilon \rightarrow 0}
[\hat{W}(x+i\epsilon) - \hat{W}(x-i\epsilon)]$. Thus for example $\rho(x)$
is given by the discontinuity across the cut in $W(z)$ and the
two point function by
\begin{eqnarray}
\langle\hat{\rho}(x)\hat{\rho}(y)\rangle &=
-{1 \over {4 \pi^2}} \lim_{\epsilon \rightarrow 0}
[\langle\hat{W}&(x+i\epsilon)\hat{W}(y+i\epsilon)\rangle
+ \langle\hat{W}(x-i\epsilon)\hat{W}(y-i\epsilon)\rangle \nonumber \\
&&- \langle\hat{W}(x+i\epsilon)\hat{W}(y-i\epsilon)\rangle
- \langle\hat{W}(x-i\epsilon)\hat{W}(y+i\epsilon)\rangle].
\label{eq:LIMIT}
\end{eqnarray}
Therefore, in view of the correspondence discussed in the previous sections,
a single mathematical quantity in the 1-matrix model
(the r.h.s. of (\ref{eq:LOOPCORR})) simultaneously gives three different
things: correlation functions of loop operators in 2-d gravity,
eigenvalue correlators in quantum chaos, and static spatial correlators in
the CSM model.

The 2-matrix model defined by (\ref{eq:Zii}) with a cubic potential
$V(A) = {1 \over 2} A^2 - g A^3$ maps onto
2-dimensional gravity coupled to the Ising model \cite{ISING}.
The two matrices $A$ and $B$ correspond to the two spin configurations
of the Ising model and the coupling constant $c$ in the action $S(A,B)$
provides the extra Boltzmann factor when adjacent spins are unequal.

\vskip 0.5cm \noindent
{\bf 7. Loop Equations and Eigenvalue Correlators}\hfill\break
Loop equations \cite{LOOPEQii,BW}\
are identities relating different correlation functions
of the loop operators. There is a hierarchy of identities through which
successively higher point correlation functions get coupled to the
lower (e.g., one or two point) correlation functions.
In the large $N$ limit, the connected higher point correlation functions
are associated with higher powers of $1/N$ as compared to the
disconnected parts; this causes the infinite hierarchy of loop
equations to break up into finite subsets of algebraic equations, from
which all correlation functions can be solved for by an inductive
procedure.

As a simple example of this procedure in the 1-matrix model consider the
identity
\begin{equation}
0=\int dA~~{\partial \over {\partial A_{ij}}}
(e^{-S} ({1 \over {z-A}})_{ij}),\label{eq:Ii}
\end{equation}
where $A$ is a real symmetric matrix, $S=N {\rm Tr} V(A)$,
$V(A) = (1/2)\mu A^2$ (gaussian orthogonal ensemble, $\beta=1$), and $i$ and
$j$
are not summed over. Expanding the
action of the derivative on its argument and summing over $i$ and $j$
one gets the identity (for details see \cite{DJS})
\begin{equation}
0=-\mu (z W(z)-1)
+{1\over 2} \langle\hat{W} (z) \hat{W} (z)\rangle
-{1\over {2N}}{\partial\over {\partial z}} W(z).\label{eq:Li}
\end{equation}
This is an exact loop equation where the one and two point correlators
are both present. Now decompose $\langle\hat{W}(z) \hat{W}(z)\rangle$
into its disconnected and connected parts:
$\langle\hat{W} (z)^2\rangle=\langle\hat{W} (z)\rangle^2
+\langle\hat{W} (z) \hat{W} (z)\rangle_c$.
In the large $N$ limit $\langle\hat{W} (z)\rangle=W(z)$ is $O(1)$
and $\langle\hat{W} (z)
\hat{W} (z)\rangle_c$ is $O({1\over N^2})$ (see below).
Then to leading order in $1/N$ we can suppress the latter as
well as the last term in (\ref{eq:Li}), obtaining the closed large $N$
loop equation $0=-\mu (z W(z)-1)  + (1/2) W(z)^2$ with the solution
$W(z)=(2 /{a^2})[z-\sqrt{z^2-a^2}]$
which has a branch cut between $z=\pm a$,
$a^2 = 2\beta/\mu$.
Using the identity $\rho(x) = (i/2\pi)\lim_{\epsilon \rightarrow 0}
[W(x+i\epsilon) - W(x-i\epsilon)]$ in the above solution
for $W(z)$, one
immediately reproduces the Wigner semicircle law (\ref{eq:DENSITY}).

To prove `large $N$  factorization'
one observes that in (\ref{eq:E}) when $g_n$ are $O(1)$,
then generically, $E \sim O(N^2)$ (e.g., for the gaussian case
when the endpoint $a=\sqrt{2\beta/\mu}$ is $O(1)$, all the eigenvalues
$\lambda_i$ are in an $O(1)$ region around the origin, and then both
the terms in (\ref{eq:E}) are $O(N^2)$). Thus to leading order
$F \equiv \ln Z$ is $O(N^2)$, say $F \approx N^2 \Gamma$, where
$\Gamma$ is a function of $g_n$ only, independent of $N$.
{}From this it follows that
$\langle {\rm Tr} A^n {\rm Tr} A^m \rangle_c =
\partial_{g_n} \partial_{g_m} \Gamma$ is $O(1)$, while the disconnected
part $\langle{\rm Tr} A^n \rangle \langle {\rm Tr} A^m \rangle$ is $O(N^2)$
since $\langle {\rm Tr} A^n \rangle = -N \partial_{g_n} \Gamma$ is $O(N)$.
Alternatively one can understand factorization geometrically.
As indicated earlier, the
$N$ dependence coming from various types of surfaces is proportional
to $N^{\chi}$, where $\chi$ is the Euler characteristic of the surface.
Let us restrict our consideration to surfaces with $h=0$ (no handles)
which will always give the leading large $N$ contributions.
$\langle\hat{W} (z) \hat{W} (z)\rangle_c$, since it contains two
insertions of the loop operator, corresponds to connected
surfaces with two boundaries whose Euler characteristic is therefore
$\chi = 2-2h-b = 0$, while the disconnected part of
$\langle\hat{W} (z)^2\rangle$, namely $\langle\hat{W} (z)\rangle^2$,
corresponds to a pair of surfaces each with one boundary (and hence each
with $\chi=1$), thereby representing a total $\chi=2$. Thus the disconnected
part is a factor $N^2$ larger than the connected part.

To obtain the $2$-point function of the loop operators, start from the
identity
\begin{equation}
0=\int dA~~{\partial \over {\partial A_{ij}}}
(e^{-S} ({1 \over {z-A}})_{ij} \hat{W}(w)).\label{eq:Iii}
\end{equation}
Expanding the action of the derivative on its argument
yields an identity connecting $2$ and $3$-point functions of the loop
operator. Expanding that identity into connected and disconnected parts
as before, one finds that terms proportional to (\ref{eq:Li}) exactly
cancel, while the connected 3-point correlator piece is suppressed by
factors of $1/N$ or $1/N^2$ compared to other terms. In the large $N$ limit
one gets the closed loop equation
\begin{equation}
0=(W(z)-\mu z)\langle \hat{W}(z) \hat{W}(w) \rangle_c +
{1 \over {N^2}}{\partial \over {\partial w}}({{W(w)-W(z)} \over {w-z}}),
\label{eq:Lii}
\end{equation}
which yields upon simplification
\begin{equation}
\langle \hat{W}(z) \hat{W}(w) \rangle_c =
{1 \over {N^2}}{a^2 \over 2\beta}{1 \over {(z-w)^2}}
{{(W(z) -W(w))}^2 \over {(1-{a^2 \over 4}W^2(z))(1-{a^2 \over 4}W^2(w))}}.
\label{eq:TWOPOINT}
\end{equation}
We have outlined the proof of this equation for the gaussian orthogonal
ensemble, $\beta=1$. The same steps repeated for the other two ensembles
yields the $\beta$ dependence displayed above. The difference enters
at the point where action of the derivative is expanded.
Concretely, for $\beta=1$,
${{\partial A_{kl}} \over {\partial A_{ij}}}=\delta_{ik}\delta_{jl}
+\delta_{il}\delta_{jk}$ whereas for $\beta=2$,
${{\partial A_{kl}} \over {\partial A_{ij}}}=\delta_{ik}\delta_{jl}$,
and a slightly more complicated expression exists for $\beta=4$.
The difference in these expressions simply reflects
the different number of independent degrees of freedom in
the three ensembles. The extension of the method of loop equations
to the $\beta=1,4$ ensembles was discussed in \cite{DJS}.
{}From (\ref{eq:TWOPOINT}) and (\ref{eq:LIMIT}) it immediately follows
that the $2$-point
connected correlator of the eigenvalue density operator is given by
(\ref{eq:TWOPOINTi}). In ref. \cite{AMJ}\ (\ref{eq:TWOPOINT}) is derived
for an arbitrary potential from the method of loop equations for $\beta=2$.
That also
establishes the universality of (\ref{eq:TWOPOINTi}) for the unitary
ensemble. We expect that
this proof of universality of (\ref{eq:TWOPOINT}) would also extend
to the other ensembles.

For the 2-matrix model, since
the two matrices are coupled, a larger number of
identities are needed to get closed equations.
A procedure similar to the one outlined above yields for the three
gaussian ensembles the global correlator \cite{DJS}
\begin{eqnarray}
\langle \hat{W}_A(z) \hat{W}_B(w) \rangle_c
&=&{1 \over N^2} {ca^2\over {2 \mu \beta}}
{1\over{(1-{ca^2\over 4 \mu }W(z)W(w))^2}}\nonumber \\
&&[{W(z)^2\over{(1-{a^2\over 4}W^2(z))}}]
[{W(w)^2\over{(1-{a^2\over 4}W^2(w))}}],\label{eq:G}
\end{eqnarray}
with $a=({{2 \beta \mu} \over {\mu^2-c^2}})^{1/2}$.
(\ref{eq:D}) follows from (\ref{eq:G}) upon using (\ref{eq:LIMIT}).
This method bypasses the problem of angular integrations.
For other applications of loop equations in hermitian, non-gaussian
2-matrix models, see \cite{LOOPEQiii}.

Finally we remark that
(\ref{eq:TWOPOINT}), (\ref{eq:TWOPOINTi}), (\ref{eq:G}), and
(\ref{eq:D}) provide `smoothed'
correlation functions because we first compute loop operator
correlation functions for $z$ and $w$ far from the cut which automatically
averages over the oscillations on the scale of the eigenvalue spacing.

\vskip 0.5cm \noindent
{\bf 8. Summary}\hfill\break
We have described some elementary mathematical properties of random
matrix models and their connection with three distinct areas:
two-dimensional quantum gravity, the $1/r^2$ integrable model, and
quantum chaos (the last connection includes applications to nuclear
physics and mesoscopic systems).
The loop operators are a link between these areas since they provide
eigenvalue correlations in quantum chaos and the CSM model
on the one hand, and also have a direct
geometric significance in quantum gravity and string theory.
The method of loop equations is a powerful method for calculating
smoothed global correlators.
This method treats all ensembles on the same footing
because the procedure for deriving the loop equations remains
the same. The only distinction between the ensembles enters
at the level of counting the degrees of freedom, that is,
in the expression for
${{\partial A_{kl}} \over {\partial A_{ij}}}$.
The extension of loop
equations to the orthogonal and symplectic ensembles should also
find application to sums over unoriented random surfaces.

\acknowledgments

I would like to thank N. Deo and B. S. Shastry for discussions and
collaboration. Part of the material discussed here was presented
in talks given at the following meetings: Winter School on Some Recent
Developments in Quantum
Many Body Physics (December 1994-January 1995, Bangalore, India);
Theoretical Physics Seminar Circuit
Discussion Meeting (April 1995, Hyderabad, India); Seventh Regional Conference
on Mathematical Physics (October 1995, Anzali, Iran). I would like to
thank the organizers of these meetings for providing a stimulating
atmosphere.

\end{document}